\documentclass[10pt]{llncs}

\usepackage[utf8]{inputenc}
\usepackage[left=4cm,right=4cm,top=4cm,bottom=4cm]{geometry}
\usepackage{wrapfig}
\usepackage{amsmath}
\usepackage{amssymb}
\usepackage{color}
\usepackage{latexsym}
\usepackage{makeidx}
\usepackage{array}
\usepackage{multicol}
\usepackage{numprint}
\usepackage{t1enc}
\usepackage{times}
\usepackage{graphicx}
\usepackage{url}
\npdecimalsign{.} 

\definecolor{mygrey}{gray}{0.75}
\newcommand{\ie}{i.e.\ }

\newcommand{\eg}{e.g.\ }

\newcommand{\hmey}[1]{}
\newcommand{\sout}[1]{}

\usepackage{amsfonts}

\newcommand{\pos}[1]{\mathbf{#1}}

\newcommand{\discussionsize}{\small}

\newcommand{\frage}[1]{}

\newdimen\endofsize\endofsize=0.5em
\def\endofbeweis{~\quad\hglue\hsize minus\hsize
                 \hbox{\vrule height \endofsize width
\endofsize}\par}

\newcount\shortyear\newcount\shorthour\newcount\shortminute
\shorthour=\time\divide\shorthour by 60\shortyear=\shorthour
\multiply\shortyear by 60\shortminute=\time\advance\shortminute by
-\shortyear
\shortyear=\year\advance\shortyear by -1900

\def\zeit{\number\shorthour:\ifnum\shortminute<10 0\number\shortminute
\else\number\shortminute\fi}

\def\rightmark{{\small \today, \zeit{}}}
\newcommand{\mytitle}{Is \emph{Nearly-linear} the same in Theory and Practice? \\ 
A Case Study with a Combinatorial Laplacian Solver}

\makeatletter
\def\fps@figure{tb}
\makeatother

\DeclareMathOperator{\im}{im}
\DeclareMathOperator{\update}{update}
\DeclareMathOperator{\query}{query}

\DeclareMathOperator{\lca}{LCA}

\DeclareMathOperator{\st}{st}
\DeclareMathOperator{\polylog}{polylog}

\newcommand{\OO}{\mathcal{O}}
\newcommand{\xopt}{x_{\text{opt}}}
\newcommand{\interval}[1][n]{[#1]}
\newcommand{\mat}{\SR^{n\times n}}

\newcommand{\SR}{\mathbb{R}}
\newcommand{\SF}{\mathbb{F}}

\newcommand{\optprob}[3]{%
    \let\mytmp\empty%
    \foreach \cond in {#3}{%
        \gappto\mytmp{& }%
        \xappto\mytmp{\expandonce{\cond}}%
        \gappto\mytmp{\\}%
    }%
    \par%
    \vspace{-.2\baselineskip}
    \begin{tabular}{@{\hspace{1em}}l>{$}l<{$}}%
    \textbf{#1} & #2\\%
    \textbf{subject to}\mytmp%
    \end{tabular}%
    \vspace{-.2\baselineskip}
    \par%
}

\newcommand{\probFont}[1]{\textsc{#1}\xspace}
\def\ie{i.\,e.\xspace}
\newcommand{\SDD}{\probFont{inv-sdd}}

\newcommand{\LPLC}{\probFont{inv-laplacian-current}}

\newcommand{\LCA}{\text{LCA}\xspace}

\usepackage{tikz,tikz-qtree}
\usetikzlibrary{svg.path,calc,decorations,circuits.ee.IEC,positioning,backgrounds,fit,shapes,arrows}
\tikzset{
    dot node/.append style={draw=black, fill=black, inner sep=0.5mm}
    every edge/.append style={>=stealth', thick},
    every path/.append style={thick},
    every arc/.append style={thick},
    every node/.append style={inner sep=0.5mm,minimum size=2.00mm,node distance=3em,thick}
}

\newdimen\XCoord
\newdimen\YCoord
\newcommand{\myplot}[1]{\includegraphics[width=.99\textwidth]{#1}}

\usepackage[ruled, vlined, linesnumbered]{algorithm2e}


\SetAlFnt{\footnotesize\sffamily}
\SetKw{KwRet}{return}
\LinesNumbered
\DontPrintSemicolon

\SetCommentSty{comsty}

\SetKwSty{kwsty}

\SetArgSty{argsty}

\SetFuncSty{funcsty}
\newcommand{\codeFont}[1]{{\small\texttt{#1}}}

\usepackage[textsize=scriptnotesize]{todonotes}
\usepackage{xspace, siunitx, rotating, mathtools, lscape, booktabs}
\usepackage{subcaption} 
\begin{document}
\title{\mytitle}
\author{Daniel Hoske\textsuperscript{1} \and Dimitar Lukarski\textsuperscript{2} \and Henning Meyerhenke\textsuperscript{1} \and Michael Wegner\textsuperscript{1}}

\date{}

\institute{\textsuperscript{1} Institute of Theoretical Informatics, Karlsruhe Institute of Technology (KIT), Karlsruhe, Germany \\ 
\textsuperscript{2} Paralution Labs UG \& Co. KG, Gaggenau, Germany}

\maketitle
\begin{abstract}
Linear system solving is one of the main workhorses in applied mathematics.
Recently, theoretical computer scientists have contributed sophisticated
algorithms for solving linear systems with symmetric diagonally dominant matrices
(a class to which Laplacian matrices belong) in 
provably nearly-linear time. While these algorithms are highly interesting from a theoretical 
perspective, there are no published results how they perform in practice.

With this paper we address this gap. 
We provide the first implementation of the combinatorial solver by
[Kelner et al., STOC 2013], which is particularly appealing for implementation due to its
conceptual simplicity. The algorithm exploits that a Laplacian matrix corresponds to a graph;
solving Laplacian linear systems amounts to finding an electrical flow in this graph with the help of 
cycles induced by a spanning tree with the low-stretch property.

The results of our comprehensive experimental study are ambivalent. They confirm a nearly-linear running time,
but for reasonable inputs the constant factors make the solver much slower than 
methods with higher asymptotic complexity. One other aspect predicted by theory
is confirmed by our findings, though: Spanning trees with lower stretch indeed reduce
the solver's running time. Yet, simple spanning tree algorithms perform in practice
better than those with a guaranteed low stretch.
\end{abstract}
\thispagestyle{empty}


\section{Introduction}
\label{chap:intro}

Solving square linear systems $Ax=b$, where $A\in \real^{n\times n}$ and $x, b \in \real^n$,
has been one of the most important problems in applied mathematics with wide applications
in science and engineering.
In practice system matrices are often \emph{sparse}, \ie they contain $o(n^2)$ nonzeros.
Direct solvers with cubic running times do not exploit sparsity. Ideally, the required time for 
solving sparse systems would grow linearly with the number of nonzeros $2m$. Moreover, 
approximate solutions usually suffice due to the imprecision of floating point arithmetic.
Spielman and Teng~\cite{Spielman2004},
following an approach proposed by Vaidya~\cite{Vaidya90}, achieved a major breakthrough in this
direction by devising a nearly-linear time algorithm for solving linear systems in
symmetric diagonally dominant matrices.
\emph{Nearly-linear} means $\OO\bigl(m\cdot\polylog(n)\cdot\log(1/\epsilon)\bigr)$ here, where $\polylog(n)$ is the set of real polynomials in $\log(n)$ and $\epsilon$ is the relative
error $\Vert x - \xopt\Vert_A / \Vert \xopt \Vert_A$ we want for the solution~$x \in \real^n$. Here $\Vert \cdot \Vert_A$ is
the norm $\Vert x \Vert_A := \sqrt{x^TAx}$ given by $A$, and $\xopt := A^{+}b$ is an exact solution.
A matrix $A = (a_{ij})_{i,j\in \interval} \in \mat$ is \emph{diagonally dominant} if $|a_{ii}| \geq \sum_{j\neq i} |a_{ij}|$ for all $i \in \interval$.
Symmetric matrices that are diagonally dominant (SDD matrices) have many
applications: In elliptic PDEs~\cite{Boman04}, maximum flows~\cite{Christiano:2011}, and sparsifying graphs~\cite{Spielman2008}. 
Thus, the problem \SDD of solving linear systems $Ax=b$ for $x$ on SDD matrices $A$
is of significant importance. We focus here on Laplacian matrices (which are SDD) due to their 
rich applications in graph algorithms, e.\,g.\ load balancing~\cite{DBLP:journals/pc/DiekmannFM99},
but this is no limitation~\cite{Kelner2013}.

\vspace{-1.75ex}
\paragraph*{Related work.}
\label{sec:related}
%
Spielman and Teng's seminal paper~\cite{Spielman2004} requires a lot of sophisticated 
machinery: a multilevel approach
\cite{Vaidya90,Reif1998} using recursive preconditioning, preconditioners based on low-stretch spanning trees~\cite{Spielman2009} and spectral graph sparsifiers \cite{Spielman2008,Koutis2012}.
Later papers extended this approach, both by making it simpler and by reducing the exponents of the polylogarithmic time factors.\footnote{Spielman provides a comprehensive
overview of later work at \url{http://www.cs.yale.edu/homes/spielman/precon/precon.html} (accessed on February 10, 2015).}
We focus on a simplified algorithm by Kelner et~al.~\cite{Kelner2013} that reinterprets the
problem of solving an SDD linear system as finding an electrical flow in a graph. It only 
needs low-stretch spanning trees and achieves 
$\OO\bigl(m\log^2\!n \log\log n \log(1/\epsilon)\bigr)$ time.

Another interesting nearly-linear time SDD solver is the 
recursive sparsification approach by Peng and Spielman~\cite{Peng2013}. Together
with a parallel sparsification algorithm, such as the one
given by Koutis~\cite{Koutis2014}, it yields a nearly-linear work
parallel algorithm.

Spielman and Teng's algorithm crucially uses the low-stretch spanning trees first introduced
by Alon et~al.~\cite{Alon95}. Elkin et~al.~\cite{Elkin2005} provide an algorithm for computing spanning trees with polynomial stretch in nearly-linear time.
Specifically, they get a spanning tree with $\OO(m\log^2\!n \log\log n)$ stretch in $O(m\log^2\!n)$ time. Abraham et~al.~\cite{Abraham2008,Abraham2012} later showed how to get rid of some of the logarithmic factors in both stretch and time.
%
%

\vspace{-1.75ex}
\paragraph*{Motivation, Outline and Contribution.}
%
Although several extensions and simplifications to Spielman and
Teng's nearly-linear time solver~\cite{Spielman2004} have been proposed, 
none of them has been validated in practice so far.
We seek to fill this gap by implementing and evaluating an algorithm proposed by Kelner et al.~\cite{Kelner2013} that is easier to describe and implement than 
Spielman and Teng's original algorithm.
%
\label{sec:outline}
%
Thus, in this paper we implement the KOSZ solver (the acronym follows from the authors' last
names) by Kelner et~al.~\cite{Kelner2013} and investigate
its practical performance. To this end, we start in Section~\ref{chap:prelim} by settling notation
and outlining KOSZ.
In Section~\ref{chap:implementation} we elaborate
on the design choices one can make when implementing KOSZ.
In particular, we explain when these choices result in a provably nearly-linear time algorithm.
Section~\ref{chap:evaluation} contains the heart of this paper, the experimental evaluation of 
the Laplacian solver KOSZ. We consider the configuration options of the algorithm,
its asymptotics, its convergence and its use as a smoother.
Our results confirm a nearly-linear running time, but at the price of very high constant factors,
in part due to memory accesses. 
We conclude the paper in Section~\ref{chap:conclusion} by summarizing the experimental 
results and discussing future research directions.

\section{Preliminaries}
\label{chap:prelim}
%
\paragraph*{Fundamentals.}
We consider undirected simple graphs $G = (V, E)$ with $n$ vertices and $m$ edges.
A graph is \emph{weighted} if we have an additional function $w\colon E \to \real_{>0}$.
Where necessary we consider unweighted graphs to be weighted with $w_e = 1 ~ \forall e \in E$.
We usually write an edge $\{u, v\} \in E$ as $uv$ and its weight as $w_{uv}$. 
Moreover, we define the set operations $\cup$, $\cap$ and $\setminus$ on graphs by
applying them to the set of vertices and the set of edges separately.
For every node $u \in V$ its \emph{neighbourhood~$N_G(u)$} is
the set $N_G(u) := \{v\in V: uv \in E\}$ of vertices $v$ with an edge to $u$ and
its \emph{degree~$d_u$} is $d_u = \sum_{v \in N_G(u)} w_{uv}$.
The \emph{Laplacian matrix} of a graph~$G = (V, E)$ is defined as
$
    L_{u,v} := -w_{uv} \text{ if $uv \in E$},
        \sum_{x \in N_G(u)} w_{ux} \text{ if $u = v$ and }
        0      \text{ otherwise}
  $  for $u, v \in V$.
%
%
A Laplacian matrix is always an SDD matrix. Another useful property of the Laplacian is 
the factorization $L = B^T R^{-1} B$,
where $B \in \real^{E\times V}$ is the \emph{incidence matrix} and $R \in \real^{E\times E}$ is
the \emph{resistance matrix} 
defined by $B_{ab,c} = 1$ if $a = c$, $= -1$ if $b=c$ and $0$ otherwise. 
$R_{e_1,e_2} = 1/w_{e_1}$ if $e_1 = e_2$ and $0$ otherwise. This holds
for all $e_1, e_2 \in E$ and $a, b, c \in V$, where we arbitrarily fix a start and end node for each edge when defining~$B$.
%
With
$
x^T L x = (Bx)^T R^{-1} (Bx) = \sum_{e \in E} (Bx)_e^2\cdot w_e \geq 0
$
(every summand is non-negative), one can see that $L$ is positive semidefinite.
(A matrix $A \in \real^{n\times n}$ is \emph{positive semidefinite} if $x^T Ax \geq 0$ for all $x \in \real^n$.)

\paragraph*{Cycles, Spanning Trees and Stretch.}
\label{sec:prelim-cycles}
A \emph{cycle} in a graph is usually defined as a simple path that returns to its starting point and a graph is called \emph{Eulerian} if there is a cycle that visits every edge exactly once.
In this work we will interpret cycles somewhat differently:
We say that a cycle in $G$ is a subgraph $C$ of~$G$ such that every vertex in $G$ is
incident to an even number of edges in $C$, \ie a cycle is a union of Eulerian graphs.
It is useful to define the addition~$C_1 \oplus C_2$ of two cycles $C_1, C_2$ to be the set of edges that occur in exactly one of the two cycles, \ie $C_1 \oplus C_2 := (C_1\setminus C_2) \cup (C_2\setminus C_1)$.
In algebraic terms we can regard a cycle as a vector $C \subseteq \SF_2^E$ such
that $\sum_{v \in N_C(u)} 1 = 0$ in $\SF_2$ for all~$u \in V$ and the cycle addition
as the usual addition on~$\SF_2^E$. We call the resulting linear space of cycles $\mathcal{C}(G)$.

In a \emph{spanning tree} (ST) $T=(V, E_T)$ of $G$ there is a unique path~$P_T(u, v)$ from every node~$u$ 
to every node~$v$. 
For any edge $e = uv \in E\setminus E_T$ (an \emph{off-tree-edge with respect to $T$}), the subgraph $e \cup P_T(u, v)$ is a cycle, the \emph{basis cycle induced by~$e$}.
One can easily show that the basis cycles form a basis of $\mathcal{C}(G)$. Thus, the basis cycles are very useful in algorithms that need to consider all the cycles of a graph.
%
%
Another notion we need is a measure of how well a spanning tree approximates the original
graph. We capture this by the \emph{stretch $\st(e) = \bigl(\sum_{e' \in P_T(u, v)} w_{e'}\bigr)/w_e$
of an edge $e = uv \in E$}. This stretch is the detour you need in order to get from one endpoint of the edge to the other if you stay in~$T$, compared to the length of the original edge.
In the literature the stretch is sometimes defined slightly differently, 
but we follow the definition in~\cite{Kelner2013} using~$w_e$.
 The \emph{stretch of the whole tree\,~$T$} is the sum of the individual stretches $\st(T) = \sum_{e\in E} \st(e)$. Finding a spanning tree with low stretch is crucial for proving the fast convergence of the KOSZ solver. 
\def\Ah{\widetilde{A}}
\def\bh{\widetilde{b}}
\def\xh{\widetilde{x}}
\def\yh{\widetilde{y}}
\paragraph*{KOSZ (Simple) Solver.}
\label{sec:solver}
%
As illustrated in Figure~\ref{fig:to-resist} in the appendix, we can regard $G$ as an electrical network 
where each edge $uv$ corresponds to a resistor with conductance $w_{uv}$ and $x$ as an
assignment of potentials to the nodes of~$G$.
Then $x_v - x_u$ is the voltage across $uv$ and $(x_v - x_u) \cdot w_{uv}$ is the resulting current 
along~$uv$. Thus, $(Lx)_u$ is the current flowing out of $u$ that we want to be equal to the 
right-hand side~$b_u$. These interpretations used by the KOSZ solver are summarized in 
Table~\ref{tbl:reinterpret} in the appendix. 
Furthermore, one can reduce solving SDD systems to the related problem
\LPLC~\cite{Kelner2013}: Given a Laplacian $L=L(G)$ and a vector $b \in \im(L)$,
compute a function $f\colon \widetilde{E} \to \real$ with 
(i) $f$ being a valid graph flow on $G$ with demand $b$ and
(ii) the potential drop along every cycle in $G$ being zero,
where a valid graph flow means that the sum of the incoming and outgoing flow at each vertex
respects the demand in $x$ and that $f(u,v) = -f(v,u) ~\forall uv \in E$. Also, 
$\widetilde{E}$ is a bidirected copy of $E$ and the potential drop of cycle $C$ is
$\sum_{e \in C} f(e) r_e$.
The idea of the algorithm is to start with any valid flow and successively adjust the flow 
such that every cycle has potential zero. We need to transform the flow back to
potentials at the end, but his can be done consistently, as all potential drops along cycles are zero.


\begin{algorithm}[t]
  \caption[\LPLC solver.]{\LPLC solver KOSZ.}\label{alg:basic2}

  \KwIn{Laplacian $L = L(G)$ and vector $b \in \im(L)$.}
  \KwOut{Solution $x$ to $Lx = b$.}

  $T \leftarrow$ a spanning tree of $G$\;
  $f \leftarrow$ unique flow with demand $b$ that is only nonzero on $T$\;
  \While{there is a cycle with potential drop $\neq 0$ in $f$}{
    $c \leftarrow$ cycle in $\mathcal{C}(T)$ chosen randomly weighted by stretch\;
    $f \leftarrow f - \frac{c^TRf}{c^TRc} c$\;
  }
  \KwRet{vector of potentials in $f$ with respect to the root of~$T$}
\end{algorithm}

Regarding the crucial question of what flow to start with and how to choose the cycle to be
repaired in each iteration, Kelner et~al.~\cite{Kelner2013} suggest using the cycle basis induced 
by a spanning tree~$T$ of~$G$ and prove that the convergence of the resulting 
solver depends on the stretch of~$T$. More specifically, they suggest starting with
a flow that is nonzero only on $T$ and weighting the basis cycles by
their stretch when sampling them. The resulting algorithm is shown as 
Algorithm~\ref{alg:basic2}; note that we may stop before all potential drops are zero and we can
consistently compute the \emph{potentials induced by~$f$} at the end by only looking at~$T$.

The solver described in Algorithm~\ref{alg:basic2} is actually just the
\codeFont{SimpleSolver} in Kelner et al.'s~\cite{Kelner2013} paper. They also show
how to improve this solver by adapting preconditioning to the setting of
electrical flows.
In informal experiments we could not determine a strategy that is consistently
better than the \codeFont{SimpleSolver}, so we do not pursue this 
scheme any further here. 
%
Eventually, Kelner et al.~\cite{Kelner2013} derive the following running time for KOSZ:
\begin{theorem}\cite[Thm.~3.2]{Kelner2013}
\label{thm:main}
\codeFont{SimpleSolver} can be implemented to run in time \linebreak 
$O(m \log^2 n \log \log n \log (\epsilon^{-1} n))$ while computing an $\epsilon$-approximation of $x$.
\end{theorem}

\section{Implementation}
\label{chap:implementation}
%
While Algorithm~\ref{alg:basic2} provides the basic idea of
the KOSZ solver, it leaves open several implementation
decisions that we elaborate on in this section.

\vspace{-1.75ex}
\paragraph*{Spanning trees.}
\label{sec:trees}
%
As suggested by the convergence result in Theorem~\ref{thm:main}, 
the KOSZ solver depends on low-stretch spanning trees.
%
Elkin et al.~\cite{Elkin2005} presented an algorithm requiring nearly-linear time 
and yielding nearly-linear average stretch. The basic idea is to recursively form a spanning tree
using a star of balls in each recursion step. We note that we use Dijkstra
with binary heaps for growing the balls and that we take care not to need
more work than necessary to grow the ball. In particular, ball growing
is output-sensitive and growing a ball $B(x, r) := \{v \in V: \text{Distance from $x$ to $v$ is $\leq r$}\}$ should require $\OO(d\log n)$ time where
$d$ is the sum of the degrees of the nodes in~$B(x, r)$.
The exponents of the logarithmic factors of the stretch of this algorithm were improved by subsequent papers 
(see Table~\ref{tbl:spanning} in the appendix), but Papp~\cite{Pap2014} showed experimentally that these improvements do
not yield better stretch in practice. In fact, his experiments suggest that the stretch of the provable algorithms is usually not better than just taking a minimum-weight spanning tree.
Therefore, we additionally use two simpler spanning trees without
stretch guarantees: A minimum-distance spanning tree with Dijkstra's algorithm and binary heaps; as well as a minimum-weight spanning
with Kruskal's algorithm using union-find with
union-by-size and path compression.

\begin{figure}[tb]
\centering
\begin{subfigure}[b]{0.35\textwidth}\centering
	\includegraphics[scale=0.75]{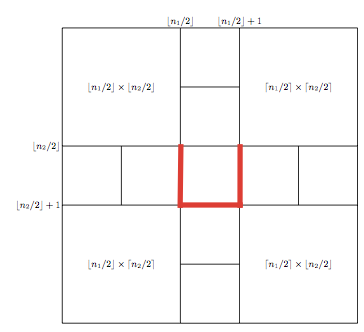}
	\caption{Recursive construction}
\end{subfigure}
\hspace{4ex}
\begin{subfigure}[b]{0.35\textwidth}\centering
	\includegraphics[scale=0.75]{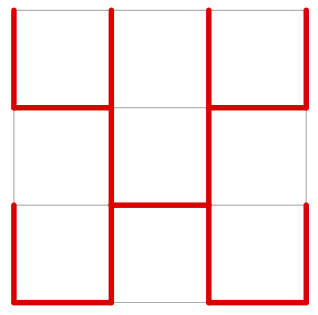}
    \caption{ST for $n_1 = n_2 = 4$}
\end{subfigure}

\caption{Special spanning tree with $\OO\bigl(\frac{(n_1+n_2)^2 \log(n_1+n_2)}{n_1 n_2}\bigr)$ average
stretch for the $n_1 \times n_2$ grid.}
\label{fig:special-rec}
\end{figure}

To test how dependent the algorithm is on the stretch of the ST, we also look at
a \emph{special ST} for $n_1 \times n_2$ grids. As depicted in 
Figure~\ref{fig:special-rec}, we construct this spanning tree by subdividing the $n_1 \times n_2$~grid into
four subgrids as evenly as possible, recursively building the STs in the
subgrids and connecting the subgrids by a U-shape in the middle.

\begin{proposition}
\label{prop:low-stretch-st}
The special ST has $\OO\bigl(\frac{(n_1+n_2)^2 \log(n_1+n_2)}{n_1 n_2}\bigr)$ average stretch on an $n_1 \times n_2$ grid.
\end{proposition}

\paragraph*{Flows on trees.}
\label{sec:flows}
%
Since every basis cycle contains exactly one off-tree-edge,
the flows on off-tree-edges can simply be stored in a single vector. 
To be able to efficiently get the potential drop
of every basis cycle and to be able to add a constant amount of flow to it,
the core problem is to efficiently store and update flows in $T$. 
More formally, we want to support the following two operations for all $u, v \in V$ and $\alpha \in \real$ on the flow~$f$:
\begin{itemize}
  \item $\query(u, v)$: return the potential drop $\sum_{e \in P_T(u, v)} f(e)r_e$
  \item $\update(u, v, \alpha)$: set $f(e) := f(e) + \alpha$ for all $e \in P_T(u, v)$
\end{itemize}

We can simplify the operations by fixing $v$ to be the root $r$ of $T$: 
$\query(u)$: return the potential drop $\sum_{e \in P_T(u, r)} f(e)r_e$ and
$\update(u, \alpha)$: set $f(e) := f(e) + \alpha$ for all $e \in P_T(u, r)$.
The itemized two-node operations can then be supported with $\query(u, v) := \query(u) - \query(v)$ and $\update(u, v, \alpha) := \bigl\{\update(u, \alpha) \text{\,and} \update(v, -\alpha)\bigr\}$
since the changes on the subpath $P_T\bigl(r, \lca(u, v)\bigr)$ cancel out. Here $\lca(u, v)$ is the \emph{lowest common ancestor} of the nodes $u$ and~$v$ in $T$, the node farthest from~$r$ that is an ancestor of both $u$ and~$v$.
We provide two approaches for implementing the operations, first an implementation 
of the one-node operations that stores the flow directly on the tree and uses the definitions of the operations without modification. Obviously, these operations require $\OO(n)$ 
worst-case time and $\OO(n)$ space. With an \LCA data structure, one can implement the 
itemized two-node operations without the subsequent simplification of using one-node operations.
This does not improve the worst-case time, but can help in practice. 
%
Secondly, we use the improved data structure by Kelner et~al.~\cite{Kelner2013} that guarantees $\OO(\log n)$ worst-case time but uses $\OO(n\log n)$ space. In this case the one-node operations boil down to a dot product ($\query$) and an addition ($\update$) of a dense vector and a sparse vector. We unroll the recursion 
within the data structure for better performance in practice.

\paragraph*{Cycle selection.}
\label{sec:selection}
%
%
The easiest way to select a cycle is to choose an off-tree edge \emph{uniformly at 
random} in $\OO(1)$~time. However, to get provably good results, we need to weight the
off-tree-edges by their stretch.
We can use the flow data structure described above to get the stretches.
More specifically, the data structure initially represents~$f = 0$. For every off-tree edge $uv$ we first execute $\update(u, v, 1)$, then $\query(u, v)$ to get $\sum_{e \in P_T(u, v)} r_e$ and finally $\update(u, v, -1)$ to return to $f = 0$.
This results in $\OO(m\log n)$~time to initialize cycle selection.
Once we have the weights, we use \emph{roulette wheel selection} in order to select a cycle in
$\OO(\log m)$~time after an additional $\OO(m)$~time initialization. 
%

%
For convenience we summarize the implementation choices for Algorithm~\ref{alg:basic2} in
Table~\ref{tbl:time} (appendix).
The top-level item in each section is the running time of the best 
sub-item that can be used to get a provably good running time. 
The convergence theorem requires a low-stretch spanning tree and weighted cycle selection.
Note that $m = \Omega(n)$ as $G$ is connected.

\section{Evaluation}
\label{chap:evaluation}
%

%
%
%
%

\subsection{Settings}
\label{sec:eval-bench}
%
We implemented the KOSZ solver in C++ using NetworKit~\cite{Staudt2014}, a toolkit
focused on scalable network analysis algorithms. As compiler we use g++ 4.8.3. The benchmark platform
is a dual-socket server with two 8-core Intel Xeon E5-2680 at 2.7 GHz each and 256 GB RAM.
Only a representative subset of our experiments are shown here. More experiments and their
detailed discussion can be found in~\cite{Hoske14experimental}. 
We compare our KOSZ implementation to
existing linear solvers as implemented by the libraries Eigen~3.2.2 \cite{Eigen} and Paralution~0.7.0~\cite{Paralution}. 
CPU performance characteristics such as 
the number of executed FLOPS (floating point operations), etc.\ are measured with the PAPI library~\cite{Browne2000}.
%
%

We mainly use two graph classes for our tests:
(i) 
Rectangular $k \times l$ grids given by $\mathbb{G}_{k,l} :=
  \bigl(\interval[k]\times\interval[l],
        \bigl\{\{(x_1, y_1), (x_2, y_2)\} \subseteq \binom{V}{2}: |x_1-x_2| = 1 \lor |y_1-y_2|=1\bigr\}\bigr)$.
	Laplacian systems on grids are, for example, crucial for solving
	boundary value problems on rectangular domains;
%
(ii) 
Barabási-Albert~\cite{Barabasi1999} random graphs with parameter $k$.
  These random graphs are parametrized with a so-called \emph{attachment~$k$}.
%
	Their construction models that the degree distribution in
	many natural graphs is not uniform at all.
%
For both classes of graphs, we consider both unweighted 
and weighted variants~(uniform random weights in $[1, 8)$).
We also did informal tests on 3D~grids and graphs that were not generated
synthetically. These graphs did not exhibit significantly different
behavior than the two graph classes above.



\subsection{Results}
\label{sec:eval-comps}
%
%
%
%

\paragraph*{Spanning tree.}
\label{sec:comp-spanning}
%
\begin{figure}[tb]
\centering
  \scalebox{0.75}{\myplot{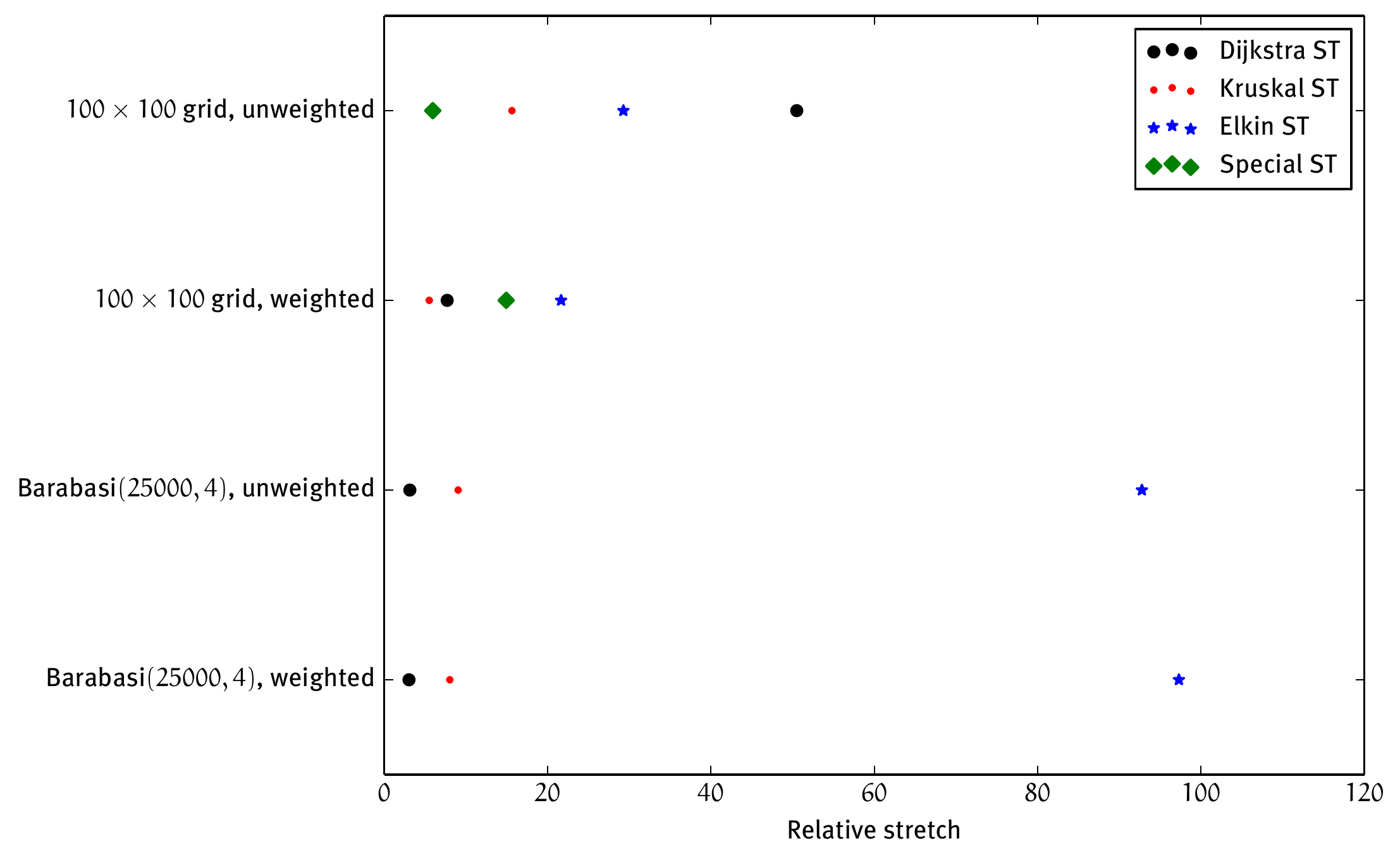}}
  \caption{Average stretch $\st(T)/m$ with different ST algorithms.}
\label{fig:stretch-bench}
\end{figure}
Papp~\cite{Pap2014} tested various low-stretch spanning tree algorithms and
found that in practice the provably good low-stretch algorithms do not yield
better stretch than simply using Kruskal. We confirm and extend this observation by
comparing our own implementation of Elkin et~al.'s~\cite{Elkin2005} low-stretch
ST algorithm to Kruskal and Dijkstra in Figure~\ref{fig:stretch-bench}.
Except for the unweighted $100\times100$ grid, Elkin has worse stretch than the
other algorithms and Kruskal yields a good ST. For Barabási-Albert graphs, Elkin
is extremely bad (almost factor~$20$ worse). Interestingly, Kruskal outperforms
the other algorithms even on the unweighted Barabási-Albert graphs, where it
degenerates to choosing an arbitrary~ST.
Figure~\ref{fig:stretch-bench} also shows that our special ST yields significantly
lower stretch for the unweighted 2D~grid, but it does not help in the weighted
case.

\paragraph*{Convergence.}
\label{sec:eval-conv}
%

\begin{figure}[tb]
\centering
      \makebox[\linewidth][c]{
          \begin{subfigure}[b]{0.5\textwidth}
            \myplot{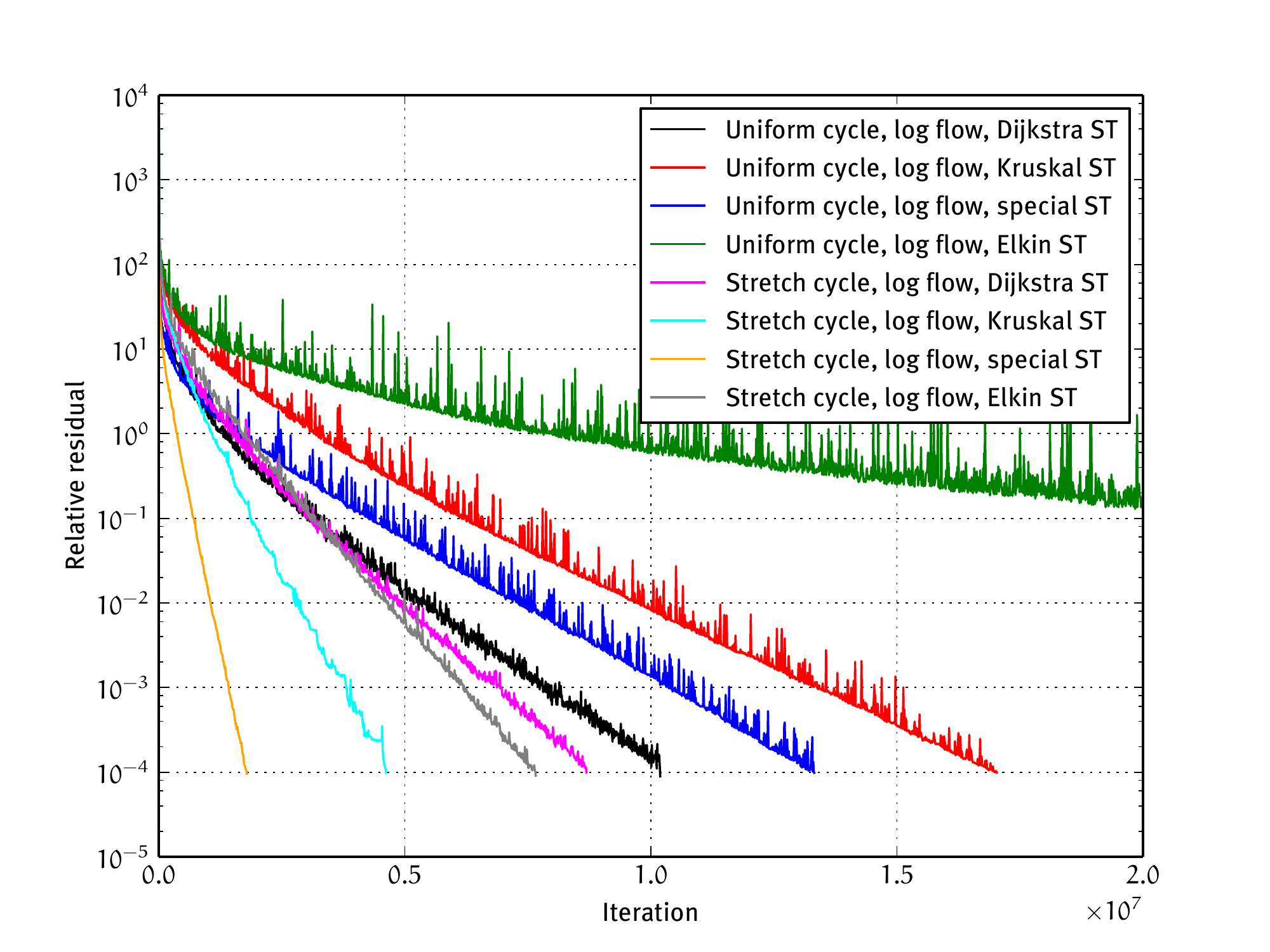}
            \caption{$100\times 100$ grid, unweighted}
          \end{subfigure}\hspace{-0.8cm}
          \begin{subfigure}[b]{0.5\textwidth}
            \myplot{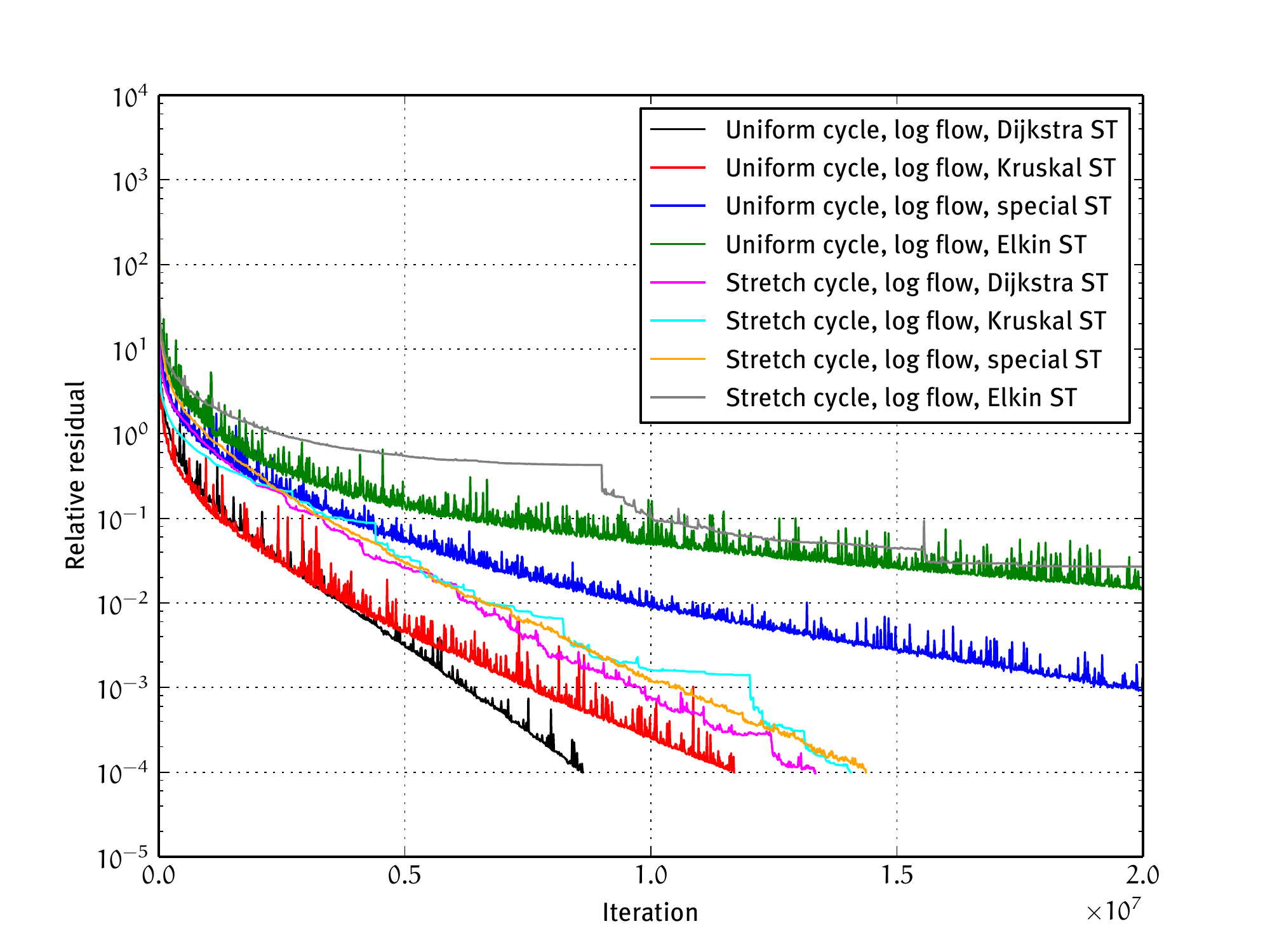}
            \caption{$100\times 100$ grid, weighted}
          \end{subfigure}
      }

      \makebox[\linewidth][c]{
          \begin{subfigure}[b]{0.5\textwidth}
            \myplot{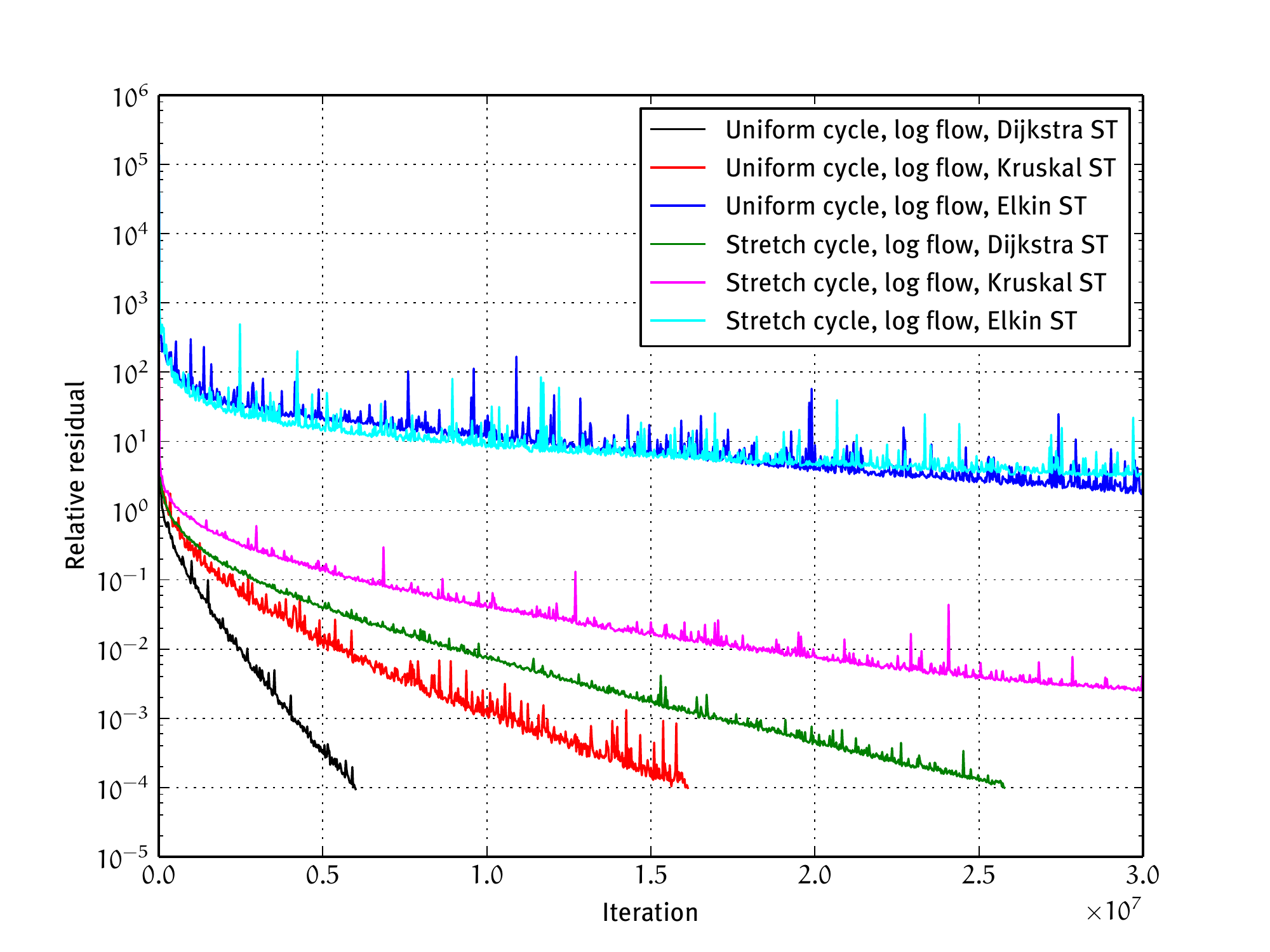}
            \caption{Barabási-Albert, $n = 25000$, unweighted}
          \end{subfigure}\hspace{-0.8cm}
          \begin{subfigure}[b]{0.5\textwidth}
            \myplot{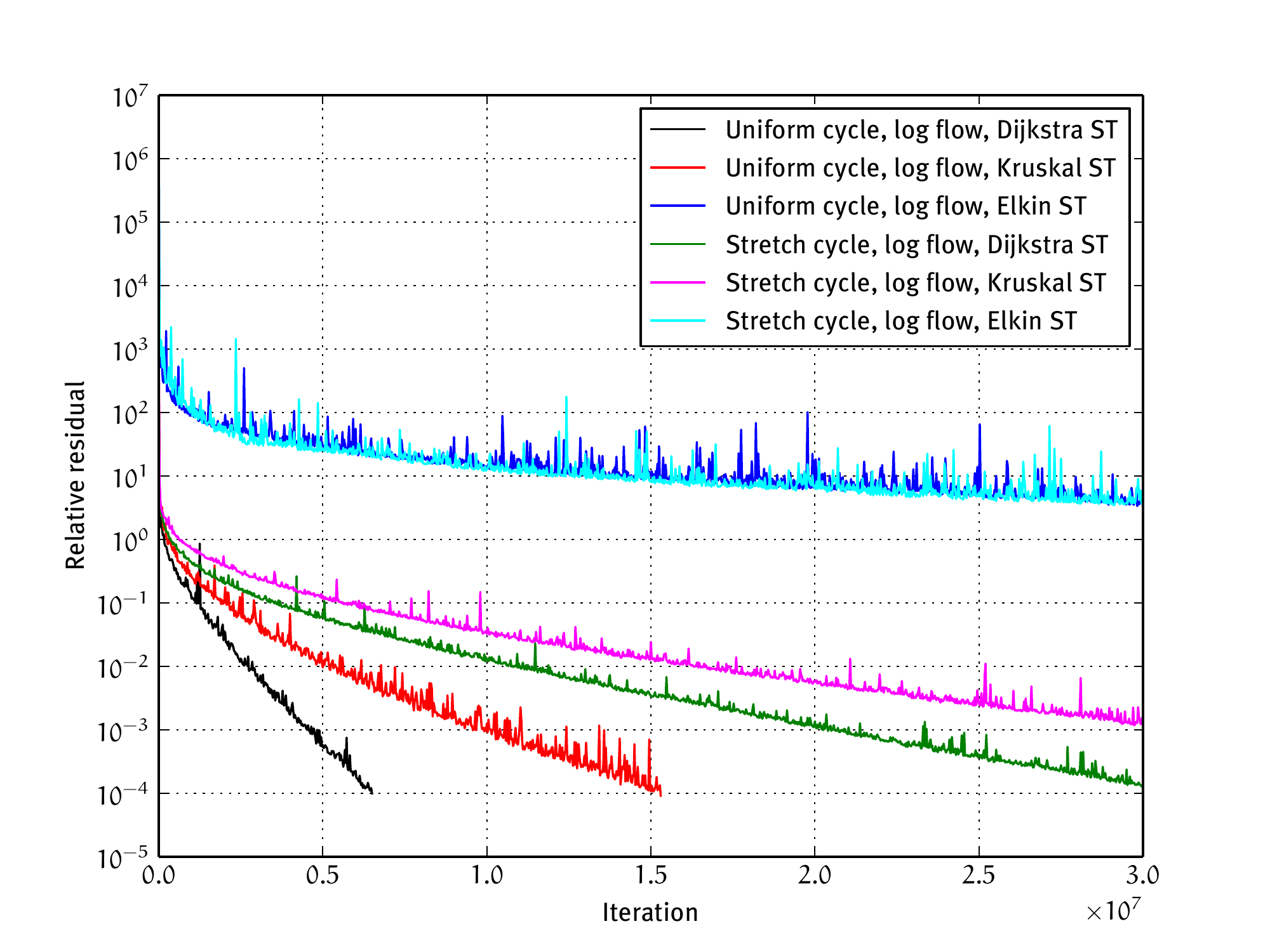}
            \caption{Barabási–Albert, $n = 25000$, weighted}
          \end{subfigure}
      }
      \caption[Convergence of the residual]{Convergence of the residual. Terminate when residual $\leq 10^{-4}$.}\label{fig:conv-residual}
\end{figure}

In Figure~\ref{fig:conv-residual} we plot the convergence
of the residual 
for different graphs and different algorithm settings. We examined a $100\times 100$~grid and a 
Barabási-Albert graph with $25,\!000$ nodes. 
%
While the residuals can increase, they follow a global downward trend. Also note that the
spikes of the residuals are smaller if the convergence is better.
%
In all cases the solver converges exponentially, but the convergence speed
crucially depends on the solver settings. If we select cycles by their stretch,
the order of the convergence speeds is the same as the order of the stretches of
the ST (cmp.\ Figure~\ref{fig:stretch-bench}), except for the Dijkstra ST and the
Kruskal ST on the weighted grid. In particular, for the Elkin ST on Barabási-Albert
 graphs, there is a significant gap to the other settings where the solver
barely converges at all and the special ST wins. Thus, low-stretch STs are
crucial for convergence. In informal experiments we also saw this behavior for
3D~grids and non-synthetic graphs.
%

We could not detect any correlation between the improvement made by a cycle repair
and the stretch of the cycle. Therefore, we cannot fully explain the different speeds with 
uniform cycle selection and stretch cycle selection. For the grid the stretch cycle selection wins, while
Barabási-Albert graphs favor uniform cycle selection.
Another interesting observation is that most of the convergence speeds stay
constant after an initial fast improvement at the start to about residual~$1$.
That is, there is no significant change of behavior or periodicity.
Even though we can hugely improve convergence by choosing the right settings,
even the best convergence is still very slow, \eg we need about $6$~million
iterations ($\approx 3000$ sparse matrix-vector
multiplications~(SpMVs) in time comparison) on a Barabási-Albert graph with
$25,\!000$~nodes and $100,\!000$~edges in order to reach residual~$10^{-4}$. In
contrast, conjugate gradient (CG)~without preconditioning only needs $204$ SpMVs for this graph.

\paragraph*{Asymptotics.}
\label{sec:eval-asymp}
%
\newcommand\largebench[2]{
  \begin{figure}[tb]
    \makebox[\linewidth][c]{
      \begin{subfigure}[b]{0.5\textwidth}
        \myplot{#1_bench_time}
        \caption{Wall time}
      \end{subfigure}\hspace{-0.8cm}
      \begin{subfigure}[b]{0.5\textwidth}
        \myplot{#1_bench_cycles}
        \caption{Cycles}
      \end{subfigure}
    }
    \makebox[\linewidth][c]{
      \begin{subfigure}[b]{0.44\textwidth}
        \myplot{#1_bench_flops}
        \caption{FLOPS}
      \end{subfigure}\hspace{-0.8cm}
      \begin{subfigure}[b]{0.5\textwidth}
        \myplot{#1_bench_memory}
        \caption{Memory accesses}
      \end{subfigure}
    }
    \caption[Asymptotics for #2]{Asymptotic behaviour for #2. Termination when relative residual was~$\leq 10^{-4}$. The error bars give the standard deviation.}
   \label{fig:bench-#1}
  \end{figure}
}
\largebench{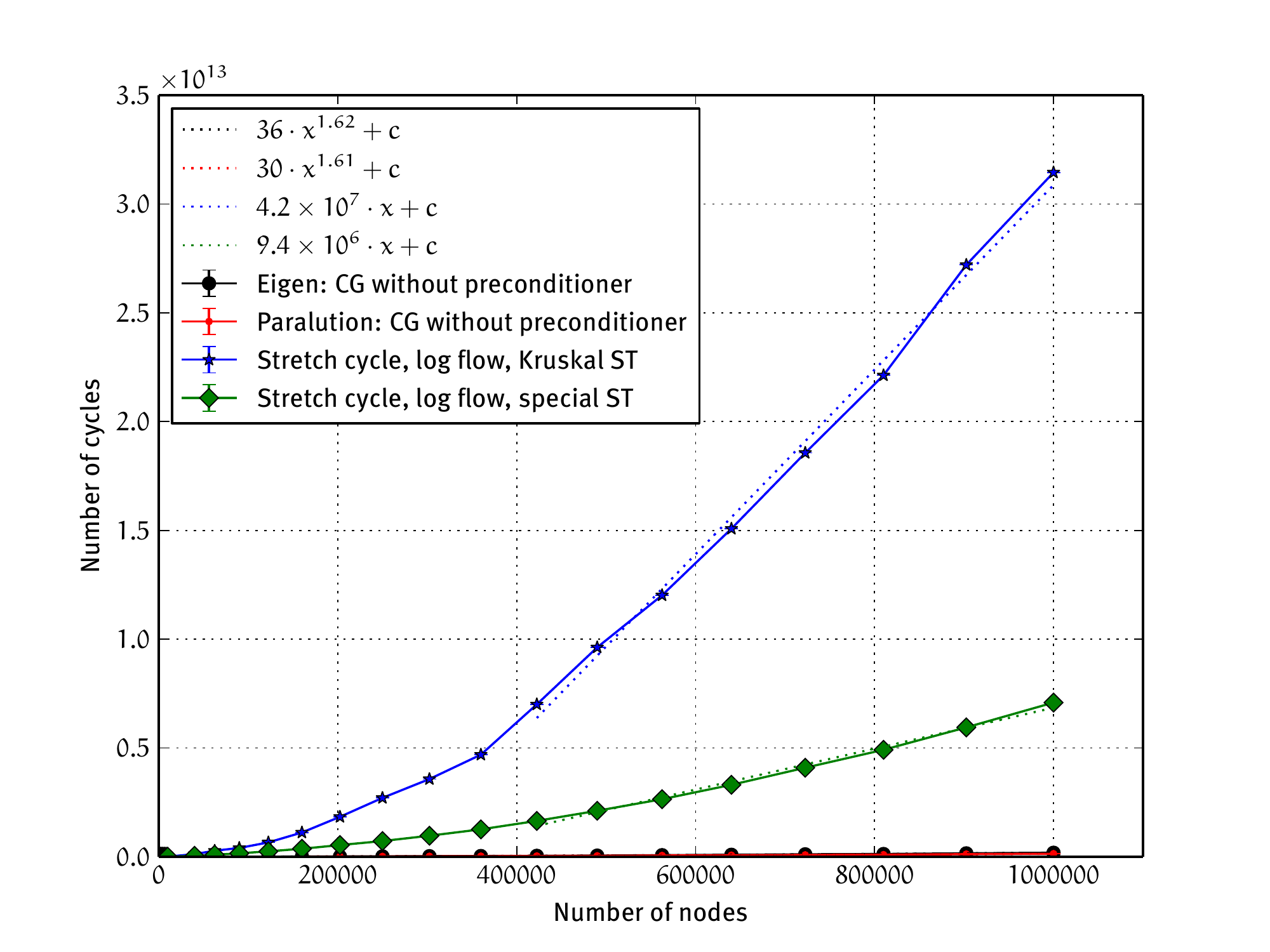}{$2D$~grids}

Now that we know which settings of the algorithm yield the best
performance for 2D~grids and Barabási-Albert graphs, we proceed by looking at how the
performance with these settings behaves asymptotically and how it compares to
conjugate gradient~(CG) without preconditioning, a simple and
popular iterative solver. Since KOSZ turns out to be not competitive, we do not need 
to compare it to more sophisticated algorithms.

In Figure~\ref{fig:bench-grid} each occurrence of $c$ stands for a new instance of a real 
constant. We expect the cost of the CG method to scale with $\OO(n^{1.5})$ on
2D~grids~\cite{Demmel1997}, while our algorithm should scale nearly-linearly.
This expectation is confirmed in the plot: Using Levenberg-Marquardt~\cite{Levenberg63} 
to approximate the curves for CG with a function of
the form $ax^b+c$, we get $b \approx 1.5$ for FLOPS and memory accesses, while
the (more technical) wall time and cycle count yield a slightly higher exponent
$b \approx 1.6$. We also see that the curves for our algorithm are almost linear
from about $650\times 650$. Unfortunately, the hidden constant factor is so
large that our algorithm cannot compete with CG even for a $1000\times
1000$~grid.
Note that the difference between the algorithms in FLOPS is significantly
smaller than the difference in memory accesses and that the difference in
running time is larger still. This suggests that the practical performance of
our algorithm is particularly bounded by memory access patterns and not by
floating point operations.
This is noteworthy when we look at our special spanning tree for the 2D~grid. We
see that using the special ST always results in performance that is better by a
constant factor. In particular, we save a lot of FLOPS (factor~$10$), while the
savings in memory accesses (factor~$2$) are a lot smaller. Even though the FLOPS
when using the special ST are within a factor of~$2$ of CG, we still
have a wide chasm in running time.
%

The results for the Barabási-Albert graphs 
are basically the same (and hence not shown in detail): 
Even though the growth is approximately linear from about
$400,\!000$~nodes, there is still a large gap between our algorithm and CG since the
constant factor is enormous. Also, the results for the number of FLOPS are again
much better than the result for the other performance counters.
In conclusion, although we have nearly-linear growth, even for
$1,\!000,\!000$~graph nodes, the KOSZ algorithm is still not competitive with CG because of
huge constant factors, in particular a large number of iterations and memory
accesses.

\paragraph*{Smoothing.}
\label{sec:eval-smoother}
%
One way of combining the good qualities of two different solvers is \emph{smoothing}. Smoothing means 
to dampen the high-frequency components of the error, which is usually done in combination with another 
solver that dampens the low-frequency error components.
It is known that in CG and most other solvers, the low-frequency components of the
error converge very fast, while the high-frequency components converge slowly.
Thus, we are interested in finding an algorithm that dampens the high-frequency
components, a good smoother. This smoother does not necessarily need to
reduce the error, it just needs to make its frequency distribution more
favorable. Smoothers are particularly often applied at each level of multigrid
or multilevel schemes~\cite{Briggs2000} that turn a good smoother into a good
solver by applying it at different levels of a matrix hierarchy.
To test whether the Laplacian solver is a good smoother, we start with a
fixed~$x$ with $Lx=b$ and add white uniform noise in $[-1, 1]$ to each of its
entries in order to get an initial vector~$x_0$.
Then we execute a few iterations of our Laplacian solver and check whether the high-frequency components of the error have been reduced. Unfortunately, we cannot directly start at the vector~$x_0$ in the solver. Our solution is to use
\emph{Richardson iteration}. That is, we transform the residual $r = b - Lx_0$ back to the source space by computing $L^{-1}r$ with the Laplacian solver, get the error $e = x - x_0 = L^{-1}r$ and then the output solution
$
 x_1 = x_0 + L^{-1}r.
$

\begin{figure}[ptb]
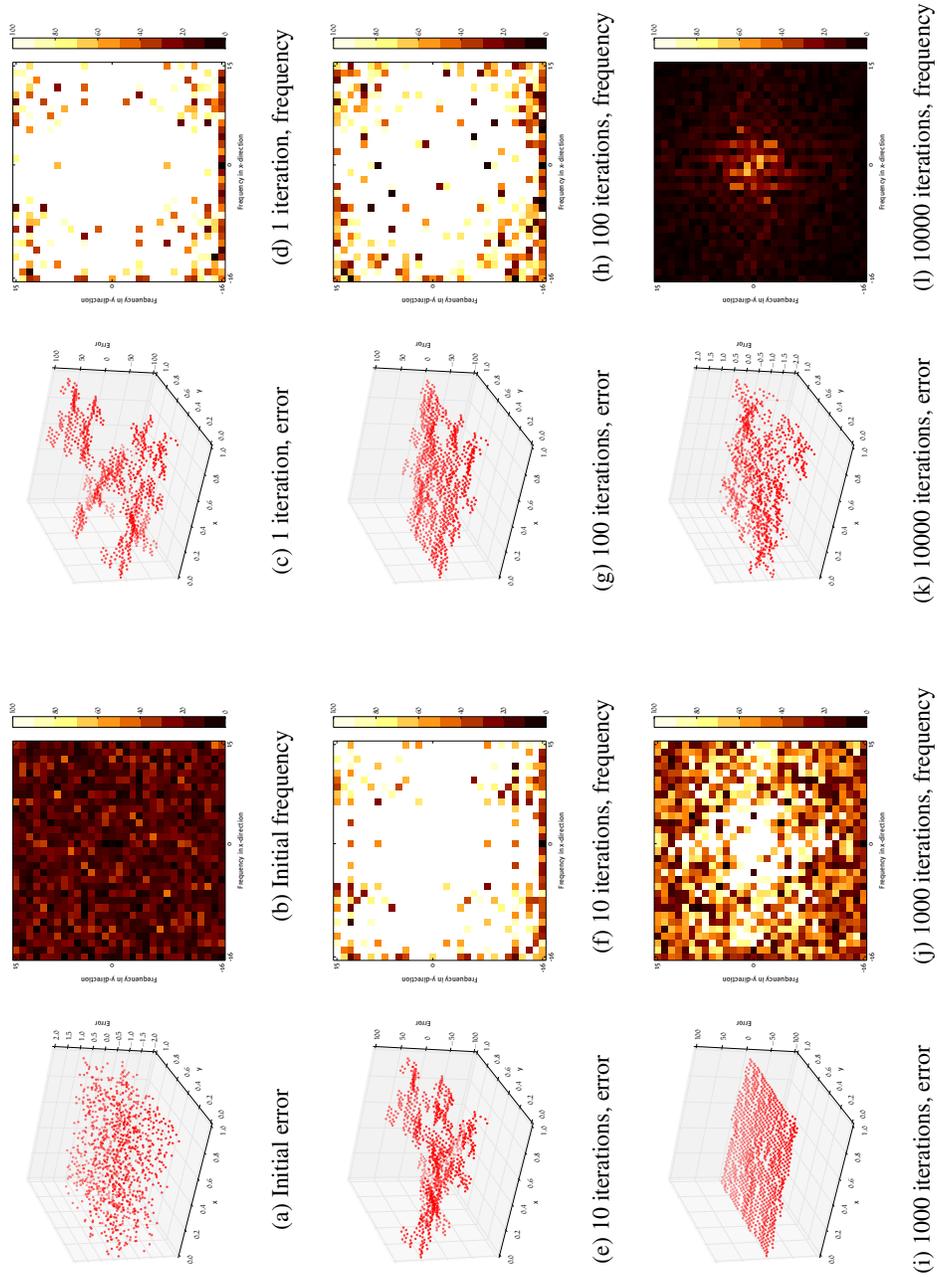
\centering
    \def\vector{smoothing_richardson_smoothing-richardson}
    \def\freq{smoothing_richardson_smoothing-freq-richardson}
    \def\plt#1#2#3{
      \begin{subfigure}[b]{0.35\textwidth}
        \myplot{\vector_#1}
        \caption{#2}
      \end{subfigure}\hspace{-0.4cm}
      \begin{subfigure}[b]{0.35\textwidth}
        \myplot{\freq_#1}
        \caption{#3}
      \end{subfigure}\hspace{-0.4cm}
    }
    \def\pltB#1{\plt{#1}{#1 iterations, error}{#1 iterations, frequency}}

      \vspace{2.5cm}
      \begin{sideways}
      \begin{minipage}{\linewidth}
      \vspace{-0.5cm}
      \makebox[\linewidth][c]{
          \plt{0}{Initial error}{Initial frequency}
          \plt{1}{1 iteration, error}{1 iteration, frequency}
      }\vspace{0.1cm}
      \makebox[\linewidth][c]{
         \pltB{10}
         \pltB{100}
      }\vspace{0.1cm}
      \makebox[\linewidth][c]{
         \pltB{1000}
         \pltB{10000}
      }
      \end{minipage}
      \end{sideways}
      \vspace{2.5cm}
      \caption[Laplacian solver as a smoother]{The Laplacian solver with the special ST as a smoother on a $32\times 32$ grid. For each number of iterations of the solver we plot the current error and the absolute values of its transformation into the frequency domain. Note that (a) and (k) have a different scale.
      }\label{fig:smoother}
      \vspace{-2\baselineskip}
\end{figure}

Figure~\ref{fig:smoother} shows the error vectors of the solver for a $32\times 32$ grid together with their transformations into the
frequency domain for different numbers of iterations of our solver. We see that the solver may indeed be
useful as a smoother since the energies for the large frequencies (on the periphery) decrease rapidly, while small frequencies (in the middle) in the error remain.

In the solver we start with a flow that is nonzero only on the ST. Therefore,
the flow values on the ST are generally larger at the start than in later
iterations, where the flow will be distributed among the other edges. Since we
construct the output vector by taking potentials on the tree, after one
iteration $x_1$ will, thus, have large entries compared to the entries of~$b$.
In subplot~(c) of Figure~\ref{fig:smoother} we see that the start vector of the solver
has the same structure as the special~ST and that its error is very large. For
the $32\times 32$~grid we, therefore, need about $10000$~iterations ($\approx
150$ SpMVs in running time comparison) to get an error of~$x_1$ similar to~$x_0$ even though the frequency
distribution is favorable. Note that the number of SpMVs the
$10000$~iterations correspond to depends on the graph size, \eg for an
$100\times100$ grid the $10000$~iterations correspond to~$20$ SpMVs.

While testing the Laplacian solver in a multigrid scheme could
be worthwhile, the bad initial vector creates robustness problems when applying the Richardson
iteration multiple times with a fixed number of iterations of our solver. In
informal tests multiple Richardson steps lead to ever increasing errors without
improved frequency behavior unless our solver already yields an almost perfect
vector in a single run.

\section{Conclusions}
\label{chap:conclusion}

At the time of writing, the presented KOSZ~\cite{Kelner2013} implementation and evaluation provide
the first comprehensive experimental study of a Laplacian solver with provably nearly-linear running
time.
Our study supports the theoretical result that the convergence of KOSZ crucially
depends on the stretch of the chosen spanning tree, with low stretch generally resulting in faster convergence.
This particularly suggests that it is crucial to build algorithms
that yield spanning trees with lower stretch. Since we have confirmd and extended 
Papp's~\cite{Pap2014} observation that algorithms with provably low stretch do not yield 
good stretch in practice, improving the low-stretch ST algorithms is an important future 
research direction.
Even though KOSZ proves to grow nearly linearly as predicted by theory, the constant seems to be
too large to make it competitive, even compared to the CG method without preconditioner.
Hence, our initial question in the paper title can be answered with ``yes'' and ``no'' at the same time:
The running time is nearly linear, but the constant factors prevent usefulness in practice.
While the negative results may predominate, our effort is the first to provide an answer at all.
We hope to deliver insights that lead to further improvements, both in theory and practice.
A promising future research direction is to repair cycles other than just the basis 
cycles in each iteration, but this would necessitate significantly different data structures.
%


\begin{small}
\bibliography{lit.bib}
\end{small}

\clearpage
\appendix
\section*{Appendix}

\section{KOSZ Solver Background}
\label{chap:apdx-solver}

\subsection{Correspondence between Graphs and Laplacian Matrices}

\begin{table}[ht]\centering
\caption[Interpretations given to the Laplacian]{Interpretations given to a Laplacian $L = L(G) \in \real^{n \times n}$ and
a vector $x \in \real^n$ where the $w_e$ for each $e \in E$ are the edge weights.}\label{tbl:reinterpret}
\smallskip
\begin{tabular}{ll}
  \toprule
  $e$   & edge/resistor $e$\\
  $w_e$ & conductance of resistor $e$\\
  $r_e := 1/w_e$ & resistance of resistor $e$\\
  $x_u$ & potential at node $u$\\
  $(Lx)_u$ & current flowing out of node $u$\\
  $b_u$ & current required to flow out of node $u$\\
  \bottomrule
\end{tabular}
\end{table}

$L$ operates on every vector $x \in \real^n$ via
\begin{align*}
  (Lx)_u & = - x_u \cdot \sum_{v \in N(u)} w_{uv} + \sum_{v \in N(u)} x_v \cdot w_{uv}\\
     & = \sum_{v \in N(u)} (x_v - x_u) \cdot w_{uv} \label{eq:lapl}
\end{align*}
for each $u \in V$.

\begin{figure}[ht]
\centering
\begin{tikzpicture}[circuit ee IEC]
\def\drawNode#1#2#3{\fill (#2) node[dot node, circle, draw=black, label=#3:$#1$] (#1) {};}
\def\drawEdge#1#2#3#4{\draw (#1) edge node[inner sep=1mm, #4] {\footnotesize $#3$} (#2);}
\def\drawGraph{
  \foreach \name/\x/\y/\pos in {1/0/0/below left,5/2/0/below right,2/1/1.73205080757/above} {
    \drawNode{\name}{\x,\y}{\pos}
  }
  \foreach \from/\to/\weight/\pos in {1/5/1/below,5/2/5/above right,1/2/2/above left} {
    \drawEdge{\from}{\to}{\weight}{\pos}
  }
}

\begin{scope}[scale=1.5]
\drawGraph
\end{scope}

\draw[dashed, ->, >=stealth'] (4, 1.5) -- (5, 1.5);

\def\drawNode#1#2#3{\fill (#2) node[dot node, circle, draw=black, label=#3:$#1V$] (#1) {};}
\def\drawEdge#1#2#3#4{\draw (#1) to[resistor] node[inner sep=2.0mm, #4] {\footnotesize $1/#3\Omega$} (#2);}
\begin{scope}[scale=1.5, xshift=4cm]
\drawGraph

\draw[very thick, ->] (0, -0.5) -- node[inner sep=1mm, below] {$(5V-1V) / 1\Omega = 4A$} (2, -0.5);
\end{scope}
\end{tikzpicture}
  \caption{Transformation into an electrical network.}\label{fig:to-resist}
\end{figure}
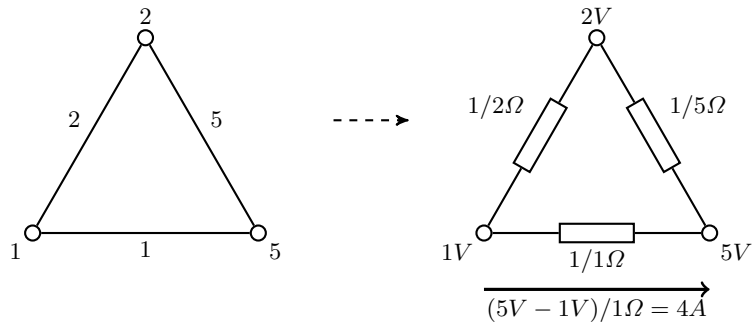

\pagebreak

\subsection{Algorithm Components}
\label{sub:components}
\begin{table}[htb]\centering
  \caption[Components of the algorithm]{Summary of the components of the algorithm}\label{tbl:time}
\smallskip
  \scalebox{0.9}{\begin{tabular}{lll}
  \toprule
  Spanning tree & $\OO\bigl(m\log n\log\log n)\bigr)$ stretch, $\OO(m \log n\log\log n)$ time \\
  \quad Dijkstra &\quad  no stretch bound, $\OO(m\log n)$ time\\
  \quad Kruskal & \quad no stretch bound, $\OO(m\log n)$ time\\
  \quad Elkin et. al. \cite{Elkin2005}& \quad $\OO(m\log^2\!n\log\log n)$ stretch, $\OO(m\log^2\!n)$ time\\
  \quad Abraham et. al. \cite{Abraham2012} & \quad $\OO(m\log n\log \log n)$ stretch, $\OO(m\log n\log\log n) $ time\\
  \midrule
  Initialize cycle selection~ & $\OO(m \log n)$ time\\
  \quad Uniform  & \quad $\OO(m)$ time\\
  \quad Weighted & \quad $\OO(m \log n)$ time\\
  \midrule
  Initialize flow & $\OO(n \log n)$ time\\
  \quad \LCA flow &  \quad $\OO(n)$ time \\
  \quad Log flow & \quad $\OO(n \log n)$ time\\
  \midrule
  Iterations & $\OO\bigl(m\log n \log \log n \log(\epsilon^{-1}\log n)\bigr)$ expected iterations\\
  \quad Select a cycle & \quad $\OO(\log n)$ time\\
  \quad\quad Uniform & \quad\quad $\OO(1)$ time\\
  \quad\quad Weighted & \quad\quad $\OO(\log n)$ time\\
  \quad Repair cycle & \quad $\OO(\log n)$ time \\
  \quad\quad \LCA flow & \quad\quad $\OO(n)$ time\\
  \quad\quad Log flow & \quad\quad $\OO(\log n)$ time\\
  \bottomrule\bottomrule\\[-2.5ex]
  Complete solver & $\OO(m\log^2\!n \log \log n \log\bigl(\epsilon^{-1}\log n)\bigr)$ expected time\\
  Improved solver & $\OO(m\log^2\!n \log \log n \log\bigl(\epsilon^{-1})\bigr)$ expected time
  \end{tabular}}
\end{table}


\section{Spanning Tree Results}
\subsection{Proof of Proposition~\ref{prop:low-stretch-st}}

We can inductively show that the average stretch $S(n_1, n_2)$ of the special ST on the $n_1 \times
n_2$ grid is in $\OO\bigl((n_1 + n_2)^2 \log(n_1 + n_2) / n_1 n_2 \bigr)$.
To do so, we first prove that by the recursive construction the distance of a
node on a border of the grid to a corner of the same border is in $\OO(n_1+n_2)$.
Thus, the stretches of the $n_1 + n_2 - 3$ off-tree edges between the rows $\lfloor
n_2/2 \rfloor$ and $\lfloor n_2/2 \rfloor + 1$ as well as the columns $\lfloor n_1/2
\rfloor$ and $\lfloor n_1/2 \rfloor + 1$ are in $\OO(n_1+n_2)$ each.
Consequently, \[S\bigl(n_1, n_2\bigr) = 4\cdot S\bigl(n_1/2, n_2/2\bigr) + \OO\bigl(n_1+n_2\bigr)^2\] when
disregarding rounding. After solving this recurrence (note that $S(n_1/2, n_2/2)$ is essentially
one fourth in size compared to $S(n_1,n_2)$), we get \[S\bigl(n_1, n_2\bigr)
= \OO\bigl((n_1 + n_2)^2 \log(n_1 + n_2)\bigr).\]
Since the number of edges of the grid is $\Theta(mn)$, the claim for the average stretch follows.
Note that in case of a square grid ($n_1 = n_2$) with $N = n_1 \times n_2$ vertices, we get
\[
S(N) = 4S(N/4) + \OO(N) = \OO(N \log N) = \OO(n_1^2 \log (n_1))
\]
and thus $\OO(\log n_1)$ average stretch.
\qed 

\pagebreak

\subsection{Overview of spanning tree algorithms and their stretch}

\begin{table}[h]
	\caption[Spanning trees and their stretch]{Spanning tree algorithms and their guaranteed stretch}\label{tbl:spanning}
	\smallskip
\centering
\small
	\begin{tabular}{lll}
		 & \textbf{Time} & \textbf{Stretch} \\
		\toprule
		\cite{Alon95} & $\OO\bigl(m^2\bigr)$ & $m\cdot exp\bigl(\OO(\sqrt{\log n \log\log n})\bigr)$ \\
		\cite{Elkin2005} & $\OO\bigl(m\log^2\!n\bigr)$ & $m \cdot \OO\bigl(\log^2 \!n\log\log n\bigr)$ \\
		\cite{Abraham2008} & $\OO\bigl(m\log^2\!n\bigr)$ & $m \cdot \OO\bigl(\log n (\log\log n)^3\bigr)$ \\
		\cite{Koutis2011} & $\OO\bigl(m\log n\log\log n\bigr)$ & $m \cdot \OO\bigl(\log n (\log\log n)^3\bigr)$\\
		\cite{Abraham2012} & $\OO\bigl(m\log n\log\log n\bigr)$ & $m \cdot \OO\bigl(\log n \log\log n\bigr)$\\
		Dijkstra~ & $\OO\bigl((m+n)\log n\bigr)$ & No guarantee\\
		Kruskal~ & $\OO\bigl(m\alpha(n)\log n\bigr)$ & No guarantee\\

		\bottomrule
	\end{tabular}
\end{table}

\end{document}